\documentclass[12pt, draftcls, onecolumn]{IEEEtran}
\usepackage{cite}

\ifCLASSINFOpdf
\else
  \usepackage[dvips]{graphicx}
\fi
\usepackage[cmex10]{amsmath}
\usepackage{array}
\usepackage[tight,footnotesize]{subfigure}
\usepackage{url}
\usepackage{amsthm}
\usepackage{amssymb}
\usepackage[nomarkers,nofiglist,notablist]{endfloat}
\theoremstyle{plain}
\newtheorem{lem}{\textbf{Lemma}}

\theoremstyle{remark}
\newtheorem*{fact}{Fact}

\begin{document}
%
\title{Design and Analysis of LT Codes with Decreasing Ripple Size}

\author{
\IEEEauthorblockN{
Jesper H. S\o rensen\IEEEauthorrefmark{1},
Petar Popovski\IEEEauthorrefmark{1},
Jan \O stergaard\IEEEauthorrefmark{1},}\\
\IEEEauthorblockA{\IEEEauthorrefmark{1}Aalborg University, Department of Electronic Systems, E-mail: \{jhs, petarp, jo\}@es.aau.dk}
}


\maketitle

\begin{abstract}
In this paper we propose a new design of LT codes, which decreases the amount of necessary overhead in comparison to existing designs. The design focuses on a parameter of the LT decoding process called the ripple size. This parameter was also a key element in the design proposed in the original work by Luby. Specifically, Luby argued that an LT code should provide a constant ripple size during decoding. In this work we show that the ripple size should decrease during decoding, in order to reduce the necessary overhead. Initially we motivate this claim by analytical results related to the redundancy within an LT code. We then propose a new design procedure, which can provide any desired achievable decreasing ripple size. The new design procedure is evaluated and compared to the current state of the art through simulations. This reveals a significant increase in performance with respect to both average overhead and error probability at any fixed overhead.
\end{abstract}
\IEEEpeerreviewmaketitle

\section{Introduction}
Rateless codes are capacity approaching erasure correcting codes. Common for all rateless codes is the ability to generate a potentially infinite amount of encoded symbols from $k$ input symbols. Decoding is possible when (1+$\alpha$)$k$ encoded symbols have been received, where $\alpha$ is close to zero. The generation of encoded symbols can be done on the fly during transmission, which means the rate of the code decreases as the transmission proceeds, as opposed to fixed rate codes, hence the name. Rateless codes are attractive due to their flexible nature. Regardless of the channel conditions, a rateless code will approach the channel capacity without the need for feedback. Moreover, practical implementations of rateless codes can be made with very low encoder and decoder complexity. The most successful examples are LT codes \cite{fc2} and Raptor codes \cite{raptor}. Originally rateless codes were intended for reliable file downloading in broadcast channels \cite{fc1}. However, lately rateless codes have drawn significant interest in the area of mobile multimedia broadcast \cite{lfc2,rcapp}.

LT codes were developed by Luby and were the first practical capacity achieving rateless code. A key part of Luby's design was the degree distribution, which is essential to a well-performing LT code. Initially Luby presented the Ideal Soliton distribution (ISD), which was shown to be optimal in terms of overhead, assuming that random processes follow expected behavior. By this we mean that when modeling the encoding and decoding processes for analysis, all random variables are assigned their expected value. Optimal behavior is achieved with the ISD, by keeping a parameter called the ripple size constantly equal to one throughout the decoding process. This parameter is described in details in section \ref{background}. A ripple size above one introduces overhead, while decoding fails if the ripple size hits zero. For this reason the ISD is optimal in theory, however, it lacks robustness against variance in the ripple size, which makes it inapplicable in practice. In order to counter this problem, Luby developed the Robust Soliton distribution (RSD), which aims at ensuring a ripple size larger than one, yet still constant. The performance of the RSD is significantly better than that of the ISD, and it is the de facto standard for LT codes. 

In \cite{lt4} the authors address the problem of finding a degree distribution, which provides a ripple of a given predefined constant size $R$. Initially, they show that such a degree distribution does not exist for high $k$, and then describe an approximate solution. In \cite{2ndmom} the variance of the ripple size is derived with the purpose of designing a robust degree distribution. The analysis is based on an assumption, which makes it valid for only ``most of the decoding process''. The authors state that their next step is to work around this assumption, in order to solve the design problem. Design criteria based on the ripple size is also applied in \cite{intermediate} and \cite{karande}. In \cite{intermediate} the goal is to find optimal degree distributions for recovery of only a fraction $0\le z < 1$ of the $k$ input symbols, while \cite{karande} aims at achieving unequal error protection across the $k$ symbols. Both papers leverage on the analytical results of \cite{hypergraphs,LTanal,hajek}.

In this work we analyze the trade-off between robustness against variance in the ripple size and required overhead (the amount of encoded symbols, in excess of $k$, necessary in order to successfully decode, i.e. $\alpha k$). Contrary to \cite{intermediate} and \cite{karande} we focus on the traditional LT code, where all data have equal importance and must be decoded. We argue that the optimal robust degree distribution for LT codes does not seek a constant ripple size. Rather a degree distribution should ensure a ripple size which decreases during the decoding process. We present a design procedure of such degree distributions and show that they outperform both the RSD and the distribution developed in \cite{lt4}.

The remainder of this paper is organized as follows. Section \ref{background} provides a brief overview of LT codes, explaining the encoding and decoding processes and relevant parameters. The analytical work of this paper is presented in section \ref{sec:analysis}, while simulation results are given in section \ref{sec:results}. Finally, conclusions are drawn in section \ref{sec:conclusion}.
\section{Background} \label{background}
\subsection{LT Codes}
In this section an overview of traditional LT codes is given. Assume we wish to transmit a given amount of data, e.g. a file or segment of a video stream. This data is divided into $k$ \textit{input symbols}. From these input symbols a potentially infinite amount of encoded symbols, also called \textit{output symbols}, are generated. Output symbols are XOR combinations of input symbols. The number of input symbols used in the XOR is referred to as the \textit{degree} of the output symbol, and all input symbols contained in an output symbol are called \textit{neighbors} of the output symbol. The output symbols of an encoder follow a certain degree distribution, $\Omega(d)$, which is a key element in the design of good LT codes. The encoding process of an LT code can be broken down into three steps:\\

\textbf{[Encoder]}
\begin{enumerate}
 \item Randomly choose a degree $d$ by sampling $\Omega(d)$.
 \item Choose uniformly at random $d$ of the $k$ input symbols.
 \item Perform bitwise XOR of the $d$ chosen input symbols. The
       resulting symbol is the output symbol.
\end{enumerate}

\noindent This process can be iterated as many times as needed, which results in a rateless code.

Decoding of an LT code is based on performing the reverse XOR operations. Initially all degree one output symbols are identified and their neighboring input symbols are moved to a storage referred to as the \textit{ripple}. Symbols in the ripple are \textit{processed} one by one, which means they are removed as content from all buffered symbols through XOR operations. Once a symbol has been processed, it is removed from the ripple and considered decoded. The processing of symbols in the ripple will potentially reduce some of the buffered symbols to degree one, in which case the neighboring input symbols are moved to the ripple. This is called a symbol \textit{release}. This makes it possible for the decoder to process symbols continuously in an iterative fashion. The iterative decoding process can be explained in two steps:\\

\textbf{[Decoder]}
\begin{enumerate}
 \item Identify all degree-1 symbols and add the neighboring input symbols to the ripple.
 \item Process a symbol from the ripple and remove it afterwards. Go to
      step 1.
\end{enumerate}

Decoding is successful when all input symbols have been recovered. If at any point before this, the ripple size equals zero, decoding has failed. This hints that a well performing LT code should ensure a high ripple size during the decoding process. However, when an output symbol is released, there is a risk that the neighboring input symbol is already in the ripple, in which case the output symbol is redundant. Hence, to minimize the risk of redundancy, the ripple size should be kept low. This trade-off was the main argument for the design goal in \cite{fc2}, that the ripple size should be kept constant at a reasonable level above one.

\section{Analysis} \label{sec:analysis}
\begin{figure}[t]
\centering
\includegraphics[width=\columnwidth]{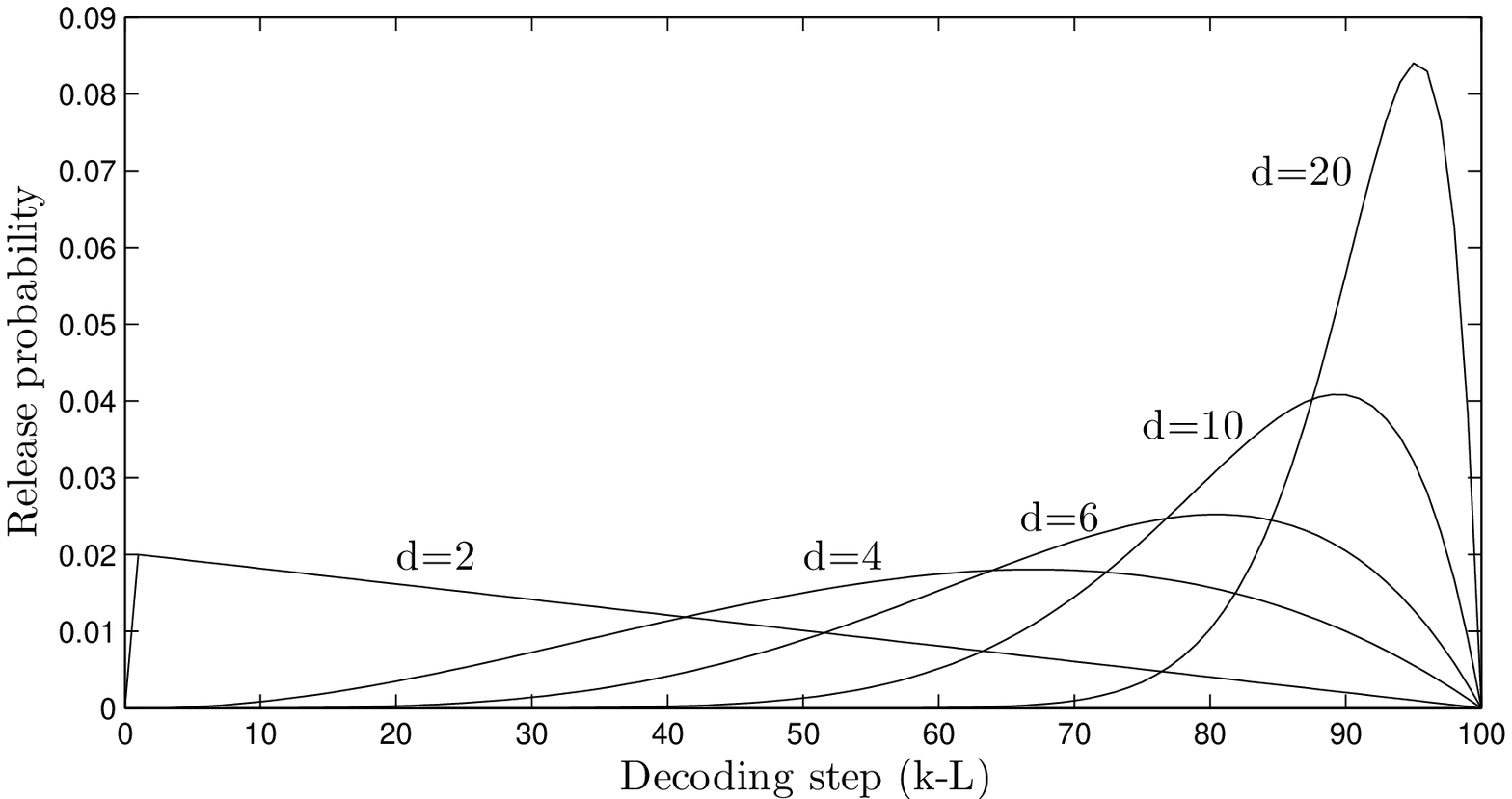}
\caption{The release probability as a function of the decoding step for fixed degrees.}
\label{fig:prop7}
\end{figure}

It is clear from the description of LT codes in section \ref{background}, that the ripple size is a very important parameter. The evolution of the ripple size is determined by the degree distribution. Thus, to obtain high decoding performance, the degree distribution should be chosen carefully, such that a desirable ripple evolution is achieved. The relation between the degree of an encoded symbol and the point of release was derived by Luby in Proposition 7 in \cite{fc2}. By point of release, we mean the point in the decoding process, where the symbol is reduced to degree one and the neighboring input symbol is potentially added to the ripple. It is parameterized by $L$, the number of remaining unprocessed symbols. The relationship is given as a probability distribution, which expresses the release probability as a function of $L$ and the original degree, $d$. Fig. \ref{fig:prop7} is a plot of the function for a number of fixed degrees, $d=2,4,6,10,20$, and $k=100$. The figure clearly shows that as the degree increases, the symbol is more likely to be released late in the decoding process, which follows intuition and the results of \cite{growth}. However, it also shows that already at quite low degrees, there is a significant probability that the symbol is not released until very late in the decoding process. 

Luby's Proposition 7 expresses the release probability only and therefore does not take into account the probability of a redundant symbol, i.e. when the achieved input symbol is already in the ripple. We show in Lemma \ref{drp}, that it is possible to take the redundancy into account and thereby express how the ripple additions are distributed.

\begin{lem}(Release and Ripple Add Probability): \label{drp}
Consider an LT code with an arbitrary degree distribution $\Omega(d)$, $d=1,...,k$. The probability, $q(d,L,R)$, that a symbol of degree $d$ is released and added to the ripple, when $L$ out of $k$ input symbols remain unprocessed, given that the ripple size is $R$ at the point of release, is

\begin{align}
 q(1,k,0) &= 1, \notag \\
 q(d,L,R) &= \frac{d(d-1)(L-R+1)\prod_{j=0}^{d-3}\left(k-(L+1)-j\right)}
 {\prod_{j=0}^{d-1}\left(k-j\right)}, \notag \\
 \textrm{for } d&=2,...,k, \notag \\
               R&=1,...,k-d+1, \notag \\
               L&=R,...,k-d+1, \notag \\
 q(d,L,R) &=0, \textrm{ for all other $d$, $L$ and $R$.} \notag
\end{align}
\end{lem}

\begin{IEEEproof}
As in the proof of Proposition 7 in \cite{fc2}, this is the probability that $d-2$ of the neighbors are among the first $k-(L+1)$ processed symbols, one neighbor is the symbol processed at step $k-L$, and the last neighbor is among the $L-R+1$ unprocessed symbols, which are not already in the ripple. This holds regardless of the choice of $\Omega(d)$. Hence,

\begin{align}
q(d,L,R) &= \frac{\binom{k-(L+1)}{d-2} \binom{1}{1} \binom{L-R+1}{1}}{\binom{k}{d}}\notag \\
         &= \frac{(L-R+1)\frac{(k-(L+1))!}{(d-2)!(k-(L+1)-(d-2))!}}{\frac{k!}{d!(k-d)!}} \notag \\
         &= \frac{(L-R+1)d!(k-d)!(k-(L+1))!}{(d-2)!k!(k-(L+1)-(d-2))!} \notag \\
         &= \frac{d(d-1)(L-R+1)\prod_{j=0}^{d-3}\left(k-(L+1)-j\right)}{\prod_{j=0}^{d-1}\left(k-j\right)}. \notag
\end{align}

\end{IEEEproof}

Using Lemma \ref{drp} we can also express the probability that a given symbol is never added to the ripple, i.e. that it is redundant. This is done in Lemma \ref{rp}.

\begin{lem}(Redundancy Probability): \label{rp}
Consider an LT code with an arbitrary degree distribution $\Omega(d)$, $d=1,...,k$. Assuming a constant ripple size $R$ during a successful decoding, i.e. where all input symbols are recovered, the probability of a symbol with degree $d$ being redundant is
\begin{align}
r(d,R) &= 1-\sum_{L=R}^{k-d+1}q(d,L,R) \notag \\
\mathrm{for} \hspace{0.2cm} d&=2,...,k-R+1, \notag \\
    R&=1,...,k-1. \notag
\end{align}
\end{lem}

\begin{IEEEproof}
When summing $q(d,L,R)$ for all $L$, we get the probability that the symbol, at some point, will be released and its neighbor added to the ripple. The remaining probability mass accounts for the events where the symbol is released, but provides an input symbol which is already in the ripple. When this happens the symbol is redundant.
\end{IEEEproof}

\begin{figure}[t]
\centering
\includegraphics[width=\columnwidth]{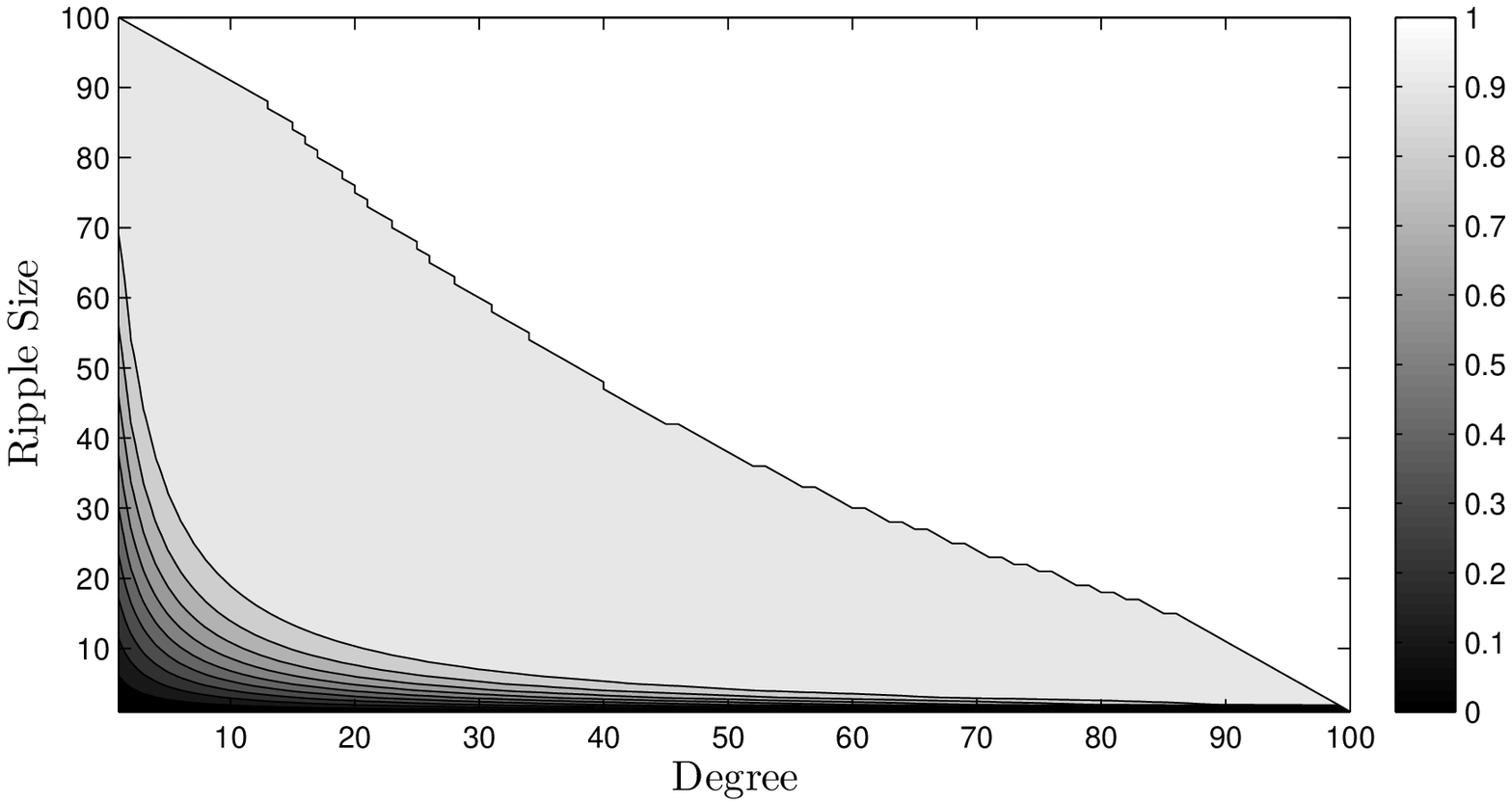}
\caption{The probability of an encoded symbol being redundant as a function of its degree and the ripple size at the point of release.}
\label{fig:colormap}
\end{figure}

Lemma \ref{rp} is quite important, since it tells us much about when redundancy occurs in an LT code. Fig. \ref{fig:colormap} shows a plot of $r(d,R)$ for $k=100$. Note that $r(d,1)=0$, $\forall$ $d$, which was expected, since a ripple size of one means that a released symbol has zero probability of providing an input symbol already being in the ripple. That is why the Ideal Soliton distribution is optimal for expected behavior. However, we must have a more robust ripple size, and even at $R$ only slightly larger than one, high degree symbols are very likely to be redundant. In general, as $d$ increases, $r(d,R)$ becomes a faster increasing function of $R$. The following fact can be deduced from Figs. \ref{fig:prop7} and \ref{fig:colormap}: 

\begin{fact}
Early in the decoding process, when mostly low degree symbols are released, a ripple size larger than one induces a relatively low probability of redundancy. Conversely, late in the decoding process, when high degree symbols are released, a ripple size larger than one induces a relatively high probability of redundancy.
\end{fact}

As mentioned in section \ref{background}, Luby sets forth a design goal of having a constant ripple size at a reasonable level above one. More specifically, he argued that the expected ripple should be kept at about $\ln(k/\delta)\sqrt{k}$, where $\delta$ is a design parameter and reflects the maximum error probability. This choice of expected ripple size evolution was motivated by the trade-off between overhead and robustness against variance in the ripple size. However, the insight gained from Lemmas \ref{drp} and \ref{rp} indicate that a well balanced trade-off is achieved with a \textit{decreasing} ripple size. A decreasing ripple size evolution is the target in the design of Raptor codes \cite{raptor}, which is a concatenation of an LT code and a fixed rate pre-code. The aim is to decode only a large fraction of the input symbols by means of the LT code, and let the pre-code provide the remaining symbol values. This code structure justifies a decreasing ripple size evolution, since a failure near the end is accounted for through a concatenated error-correcting code. However, we argue that even without the pre-code, i.e. for an LT code only, the ripple size should decrease during the decoding process, since this provides a better trade-off between overhead and robustness against ripple variance. How the ripple size should decrease during decoding is analyzed in the following subsection.

\subsection{Choosing a Ripple Evolution}
In the design of Raptor codes, the choice of ripple size evolution was based on a random walk model of the ripple \cite{raptor}. It was assumed that the ripple size either increases with one or decreases with one, with equal probabilities, in each decoding step. Thus, 

\begin{gather}\label{srw}
 R(L-1) =
 \begin{cases}
  R(L)+1, &\textrm{w. prob. $\frac{1}{2}$}, \\
  R(L)-1, &\textrm{w. prob. $\frac{1}{2}$}, \\
 \end{cases}
\end{gather}
where $R(L)$ is the ripple size when $L$ input symbols remain unprocessed. This model is the simple symmetric one-dimensional random walk model, for which the theory is well established \cite{rw}. One element is the second moment, $\overline{\Delta^2}$, where $\Delta$ is the position relative to the origin. The second moment grows linearly with the number of steps in the walk, $N$, such that $\overline{\Delta_N^2}=N$ (see \cite{rw}, page 6). The second moment quantifies the variance of the ending point of the random walk, and is thus particularly interesting in the context of LT codes. When $L$ input symbols remain unprocessed, $L$ steps remain in the random walk, and thus $\overline{\Delta_L^2}=L$, where $\Delta_L=R(0)-R(L)$. In \cite{raptor} it was heuristically argued that the expected ripple size should be kept in the order of the root mean square (RMS) distance from the origin, defined as $\sqrt{\overline{\Delta_L^2}}=\sqrt{L}$. Hence, the target ripple size evolution for Raptor codes is $R(L)=c\sqrt{L}$ for $k\ge L \ge 0$, where $c$ is a suitably chosen constant. This is an example of a decreasing ripple and is thus a candidate for our design of an LT code. However, we will now show that ripple size evolutions with exponents of $L$ lower than $\frac{1}{2}$ should be considered.

The random walk model applied in the design of Raptor codes is simplified and heuristic, which the author itself is the first to mention. There are several phenomena in the LT decoder, which are not accounted for, such as additions of multiple symbols in a single step and absorbing barriers at $R(L)=L$ and $R(L)=0$. It is outside the scope of this paper to create an accurate random walk model of the ripple. However, there is one particular phenomena which greatly influences how the ripple size evolves, and that is the bias towards zero. As decoding progresses, the ripple size will inevitably end up equal to zero, it is just a matter of when. The bias is explained by the increasing probability that a new recovered symbol is already in the ripple, $p_r=\frac{R(L+1)-1}{L}$, where $R(L+1)$ is the ripple size after the previous decoding step. We modify the random walk model in order to take this into account.

\begin{figure}[t]
\centering
\includegraphics[width=\columnwidth]{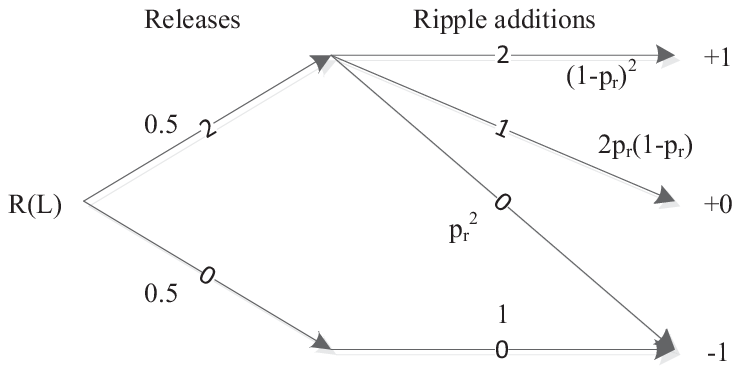}
\caption{An event tree illustrating the possible numbers of releases and additions to the ripple in a single decoding step.}
\label{fig:events}
\end{figure}

Consider the event tree in Figure \ref{fig:events}. It illustrates possible numbers of released symbols and corresponding possible numbers of symbols added to the ripple in a decoding step. These are indicated by the values breaking the line of the arrows. As an example, an arrow broken by a zero in the column named "Releases", refers to the event where the processing of a symbol results in no new symbols being released. Similarly, the arrow broken by a zero under "Ripple additions", refers to the event where none of the released symbols provide an input symbol which is added to the ripple. The value next to an arrow is the probability of the event. The event tree is constructed based on the simplifying assumptions behind \eqref{srw}, with the modification that redundancy is taken into account. This limits the possible numbers of released symbols to $0$ or $2$ and possible numbers of symbols added to the ripple to $0$, $1$ or $2$. If we disregard the probability of redundancy, i.e. $p_r=0$, the event tree reduces to the simple random walk in \eqref{srw}. However, if we take a non-zero $p_r$ into account, then probability mass is moved from outcome $+1$ to outcomes $+0$ and $-1$, which provides the bias towards a ripple size of zero. In order to ensure an initial bias of zero, we only take the increase in redundancy probability into account, i.e. $p_r'=p_r-p_{r0}$, where $p_{r0}$ is the probability of redundancy at $k-L=0$. We now arrive at the following random walk model:

\begin{gather}\label{mrw}
 R(L-1) =
 \begin{cases}
  R(L)+1, &\textrm{w. prob. $\frac{1}{2}(1-p_r')^2$}, \\
  R(L)  , &\textrm{w. prob. $p_r'(1-p_r')$}, \\
  R(L)-1, &\textrm{w. prob. $\frac{1}{2}+\frac{1}{2}p_r'^2$}. 
  \end{cases}
\end{gather}

\begin{figure}[t]
\centering
\includegraphics[width=\columnwidth]{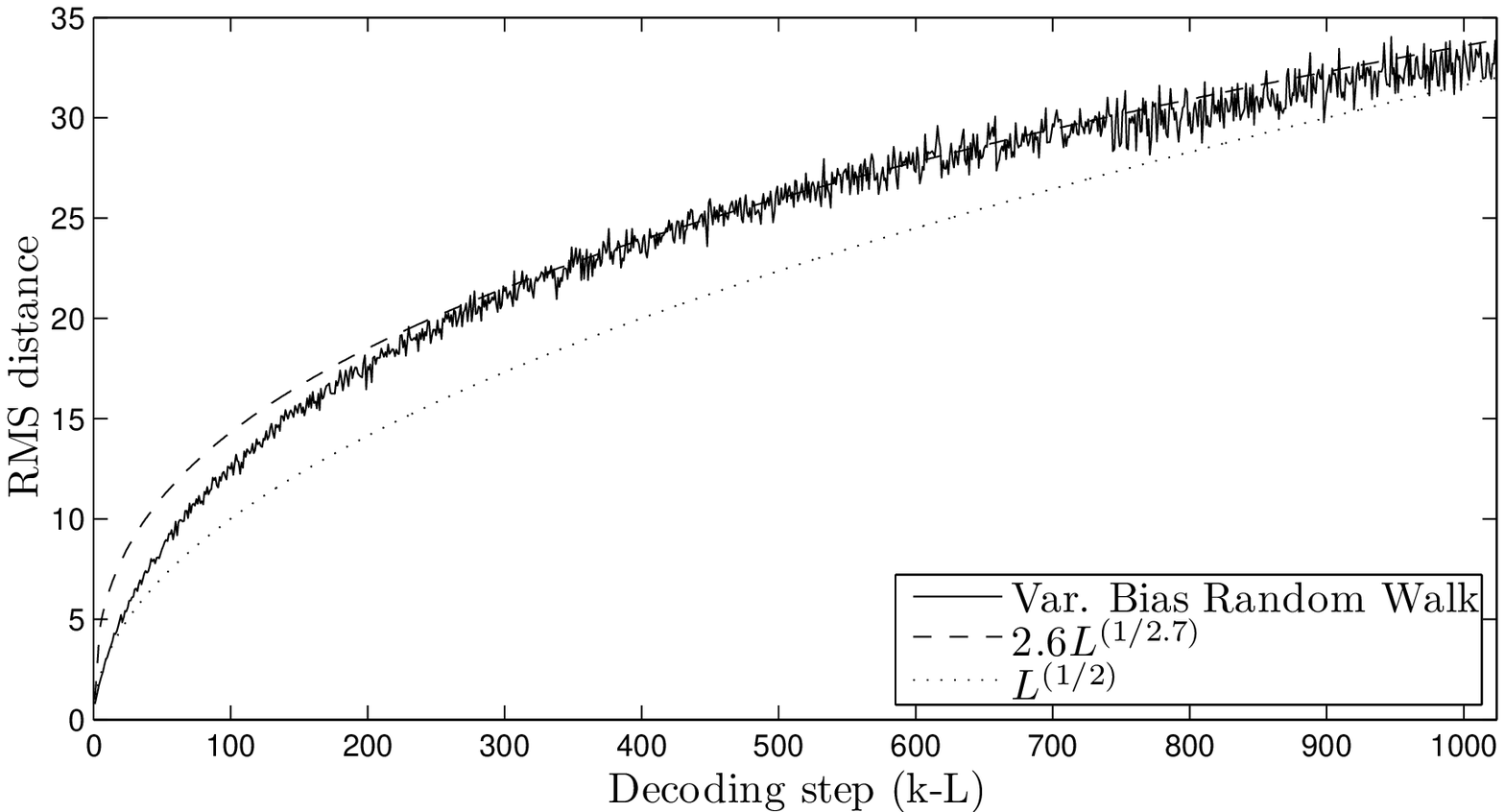}
\caption{Simulation of the proposed random walk model compared to selected relationships with $L$.}
\label{fig:randwalk}
\end{figure}

The random walk in \eqref{mrw} has a variable bias. Such a random walk is complicated to analyze, thus to find the RMS distance from the origin as a function of $L$, we use a Monte Carlo simulation. The result is shown in Figure \ref{fig:randwalk}. It is evident that the square root relationship with $L$ is no longer accurate. The figure suggests that $\sqrt{\overline{\Delta_L^2}}=c_1 L ^{(1/c_2)}$, where $c_1>0$ and $c_2>2$, is better able to capture the dynamics of the RMS distance. Hence, in this paper we will investigate ripple size evolutions of the following form:

\begin{align}
&R(L)=c_1 L ^{(1/c_2)} &\textrm{ if $c_1 L ^{(1/c_2)}\le L$},\notag \\
&R(L)=L &\textrm{ if $c_1 L ^{(1/c_2)} > L$},\notag \\
\textrm{for}& \textrm{ suitably chosen } c_1>0 \textrm{ and } c_2 \ge 2.
\end{align}

\subsection{Achieving the Desired Ripple}
Now that we have chosen a target expected ripple size evolution, it remains to be shown how to achieve this target. If $R(L)$ is the desired ripple size when $L$ symbols remain unprocessed, then the expected amount of symbols that must be added to the ripple in the $(k-L)$'th decoding step is


\begin{align}\label{target}
Q(L)&=R(L), &\textrm{for $L=k$,} \notag \\
Q(L)&=R(L)-R(L+1)+1, &\textrm{for $k>L\geq 0$.}
\end{align}

The achieved $Q(L)$ depends on the degree distribution and is expressed in \eqref{ach_Q} using Lemma \ref{drp}. 

\begin{align}\label{ach_Q}
 Q(L) &= \sum_{d=1}^{k} n \Omega(d) q(d,L,R(L+1)),
\end{align}
where $n$ is the number of received symbols included in the decoding. \eqref{ach_Q} is based on the assumption that all releases in a single step are unique, i.e. the same input symbol will not be recovered twice within a single step. This is a valid assumption, since the expected number of releases in a single step is small compared to $L$.

If $R(k+1)$ is defined as zero (ripple size before decoding starts), the following matrix equation can be constructed:

\begin{align}\label{mateq}
 \begin{bmatrix}
  q(1,k,R(k+1)) & 0 & 0 \\
  \vdots  &  \ddots & 0 \\
  q(1,1,R(2)) & \cdots & q(k,1,R(2)) \\
 \end{bmatrix} 
  \begin{bmatrix}
  n\Omega(1)  \\
  \vdots   \\
  n\Omega(k)  \\
 \end{bmatrix} &=
  \begin{bmatrix}
  Q(k)  \\
  \vdots   \\
  Q(1)
 \end{bmatrix}
\end{align}

\noindent where the LHS is based on \eqref{ach_Q} and the RHS is the target in \eqref{target}. Note that this equation provides a solution to how the $n$ received symbols should be distributed across the possible degrees. Hence, since $n$ is free, it acts as the normalization factor, in order to arrive at a valid degree distribution, as well as a measure of how many symbols must be collected in order to achieve the target $R(L)$. This matrix equation is singular for high $k$. However, finding a least squares nonnegative solution provides a ripple evolution very close to the target. Table \ref{degdist1} and Table \ref{degdist2} show solutions for $R(L)=1.7L^{1/2.5}$ at $k=256$ and $R(L)=1.9L^{1/2.6}$ at $k=1024$, respectively. These solutions have error vectors with squared norms of $0.0011$ and $0.0048$, respectively, which are negligible. The procedure explained in this subsection is a simple approach to the design of a degree distribution having any achievable expected ripple size evolution.

\begin{table}[ht]
\begin{minipage}[b]{0.48\columnwidth}\centering
\caption{Degree distribution for $R(L)=1.7L^{1/2.5}$ at $k=256$.}
\label{degdist1}
\begin{tabular}{|c|c|}
  \hline                       
  $d$ & $\Omega(d)$ \\ \hline \hline
    1 &   0.0534 \\ \hline
    2 &   0.4530 \\ \hline
    3 &   0.1538 \\ \hline
    4 &   0.0784 \\ \hline
    5 &   0.0542 \\ \hline
    7 &   0.0750 \\ \hline
   12 &   0.0392 \\ \hline
   13 &   0.0200 \\ \hline
   25 &   0.0266 \\ \hline
   26 &   0.0090 \\ \hline
   57 &   0.0152 \\ \hline
   58 &   0.0057 \\ \hline
  138 &   0.0067 \\ \hline
  139 &   0.0098 \\ \hline
  \end{tabular}\end{minipage}
\hspace{0.5cm}
\begin{minipage}[b]{0.48\columnwidth}
\centering
\caption{Degree distribution for $R(L)=1.9L^{1/2.6}$ at $k=1024$.}
\label{degdist2}
\begin{tabular}{|c|c|}
  \hline                       
  $d$ & $\Omega(d)$ \\ \hline \hline
    1  &  0.0250 \\ \hline
    2  &  0.4750 \\ \hline
    3  &  0.1600 \\ \hline
    4  &  0.0784 \\ \hline
    5  &  0.0605 \\ \hline
    7  &  0.0633 \\ \hline
    8  &  0.0109 \\ \hline
   12  &  0.0516 \\ \hline
   13  &  0.0003 \\ \hline
   22  &  0.0229 \\ \hline
   23  &  0.0097 \\ \hline
   45  &  0.0163 \\ \hline
   46  &  0.0024 \\ \hline
   98  &  0.0001 \\ \hline
   99  &  0.0104 \\ \hline
  236  &  0.0021 \\ \hline
  237  &  0.0043 \\ \hline
  601  &  0.0012 \\ \hline
  602  &  0.0057 \\ \hline
\end{tabular}\end{minipage}
\end{table}
\section{Numerical Results} \label{sec:results}
In this section, the performance of the proposed degree distribution, $\Omega(d)$, is simulated and compared to the RSD and the distribution proposed in \cite{lt4}, denoted $\beta(d)$. We will compare the average overhead required for successful decoding, i.e. when all $k$ input symbols have been recovered. Moreover, we evaluate the block error rate as a function of the overhead in order to show a more complete picture of the performance of the distributions. The distributions are simulated at $k=\{256,512,1024,2048\}$.

Initially, we present a numerical optimization of the design parameters for $\Omega(d)$, $c_1$ and $c_2$, and the RSD, $c$ and $\delta$. The optimization is performed with respect to the average overhead through $5000$ iterations. Table \ref{tab:omega} and Table \ref{tab:rsd} show the results for $k=1024$ and a selected set of combinations of the design parameters. Similarly, we have optimized the parameters for $k=256,512$ and $2048$. Optimized $\Omega(d)$ for $k=256$ and $k=1024$ are shown in Table \ref{degdist1} and Table \ref{degdist2}, respectively. It should be noted, that a design of the RSD with $\delta \ge 1$ seems not very reasonable, given the interpretation of that parameter as the maximum error probability. However, the analysis for finding the bound of the error probability in \cite{fc2} is very conservative and therefore the interpretation is not entirely accurate. Thus, $\delta \ge 1$ should still be considered, and as the results show, it provides the better performance on average.

\begin{table}[p]
\centering
\caption{Avg. overhead $(1+\alpha)$ of $\Omega$ for $k=1024$ and varying $c_1$ and $c_2$.}
\label{tab:omega}
\begin{tabular}{|c|c|c|c|c|c|c|c|c|c|c|c|}
  \hline                       
$c_1$\textbackslash$c_2$ &   2.2   &   2.3   &   2.4   &   2.5   &   2.6   &   2.7   &   2.8  \\ \hline \hline
    	1.0 	  					 &  1.131  &  1.129  &  1.129  &  1.132  &  1.130  &  1.133  &  1.133 \\ \hline
    	1.1								 &  1.126  &  1.123  &  1.122  &  1.124  &  1.126  &  1.127  &  1.125 \\ \hline
    	1.2								 &  1.117  &  1.118  &  1.117  &  1.118  &  1.119  &  1.120  &  1.120 \\ \hline
    	1.3								 &  1.109  &  1.111  &  1.109  &  1.110  &  1.115  &  1.114  &  1.114 \\ \hline
    	1.4								 &  1.099  &  1.100  &  1.102  &  1.102  &  1.105  &  1.105  &  1.105 \\ \hline
    	1.5								 &  1.093  &  1.092  &  1.094  &  1.093  &  1.094  &  1.097  &  1.099 \\ \hline
    	1.6								 &  1.092  &  1.092  &  1.091  &  1.091  &  1.092  &  1.092  &  1.093 \\ \hline
    	1.7								 &  1.090  &  1.089  &  1.090  &  1.091  &  1.090  &  1.090  &  1.092 \\ \hline
    	1.8								 &  1.093  &  1.089  &  1.087  &  1.088  &  1.087  &  1.089  &  1.090 \\ \hline
    	1.9								 &  1.092  &  1.094  &  1.126  &  1.088  &  \textbf{1.087}  &  1.088  &  1.089 \\ \hline
    	2.0								 &  1.093  &  1.090  &  1.090  &  1.091  &  1.095  &  1.127  &  1.091 \\ \hline
    	2.1								 &  1.105  &  1.094  &  1.090  &  1.090  &  1.090  &  1.090  &  1.091 \\ \hline
    	2.2								 &  1.109  &  1.284  &  1.101  &  1.093  &  1.090  &  1.089  &  1.089 \\ \hline
\end{tabular}
\end{table}

\begin{table}[p]
\centering
\caption{Avg. overhead $(1+\alpha)$ of RSD for $k=1024$ and varying $c$ and $\delta$.}
\label{tab:rsd}
\begin{tabular}{|c|c|c|c|c|c|c|c|c|c|c|c|}
  \hline                       
$c$\textbackslash$\delta$&   0.5   &   1.0   &   1.5   &   2.0   &   3.0   &   4.0   &   5.0  \\ \hline \hline
      0.01               &  1.197  &  1.207  &  1.213  &  1.224  &  1.236  &  1.251  &  1.259 \\ \hline
      0.02               &  1.139  &  1.143  &  1.146  &  1.152  &  1.158  &  1.168  &  1.172 \\ \hline
      0.03               &  1.126  &  1.126  &  1.126  &  1.127  &  1.128  &  1.136  &  1.140 \\ \hline
      0.04               &  1.126  &  1.121  &  1.118  &  1.118  &  1.117  &  1.120  &  1.125 \\ \hline
      0.05               &  1.132  &  1.121  &  1.118  &  1.117  &  1.115  &  1.114  &  1.116 \\ \hline
      0.06               &  1.139  &  1.126  &  1.119  &  1.116  &  1.112  &  1.111  &  1.112 \\ \hline
      0.07               &  1.145  &  1.130  &  1.123  &  1.118  &  1.114  &  \textbf{1.111}  &  1.111 \\ \hline
      0.08               &  1.155  &  1.138  &  1.129  &  1.121  &  1.115  &  1.113  &  1.112 \\ \hline
      0.09               &  1.165  &  1.145  &  1.133  &  1.127  &  1.119  &  1.115  &  1.113 \\ \hline
      0.10               &  1.174  &  1.151  &  1.139  &  1.132  &  1.122  &  1.118  &  1.114 \\ \hline
\end{tabular}
\end{table}

\begin{figure}[t]
\centering
\includegraphics[width=\columnwidth]{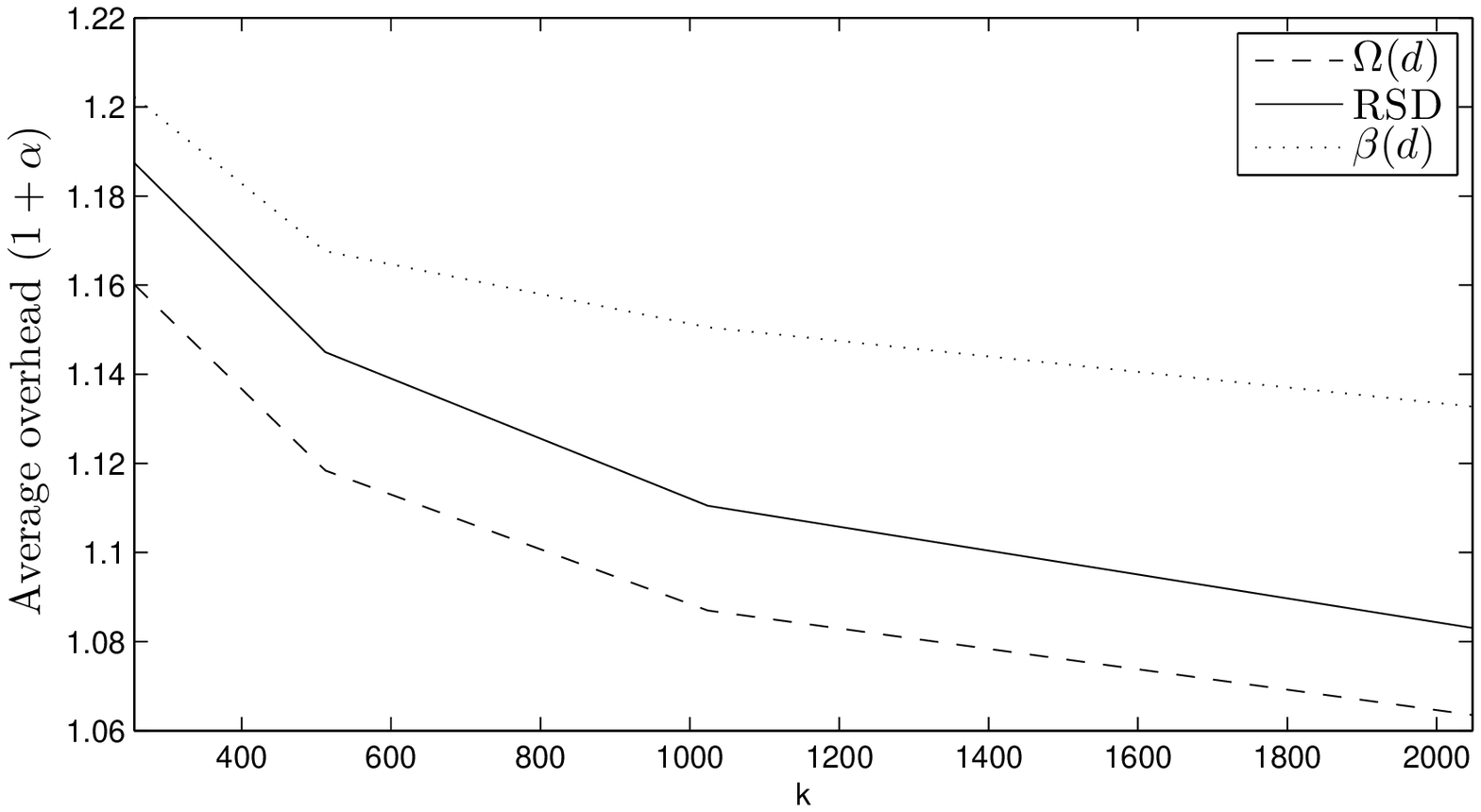}
\caption{Comparison of optimized degree distributions with respect to average overhead.}
\label{fig:avg_comp}
\end{figure}

\begin{figure}[t]
\centering
\includegraphics[width=\columnwidth]{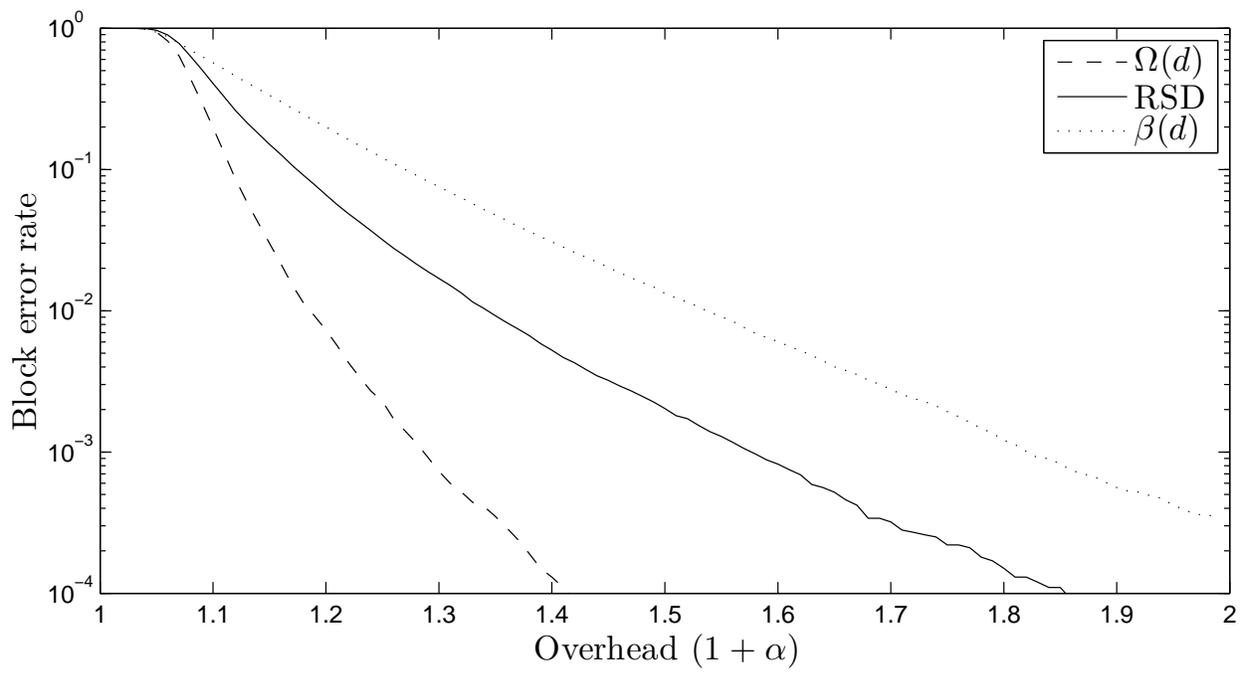}
\caption{Comparison of optimized degree distributions with respect to block error rate.}
\label{fig:1024_comp}
\end{figure}

\begin{figure}[t]
\centering
\includegraphics[width=\columnwidth]{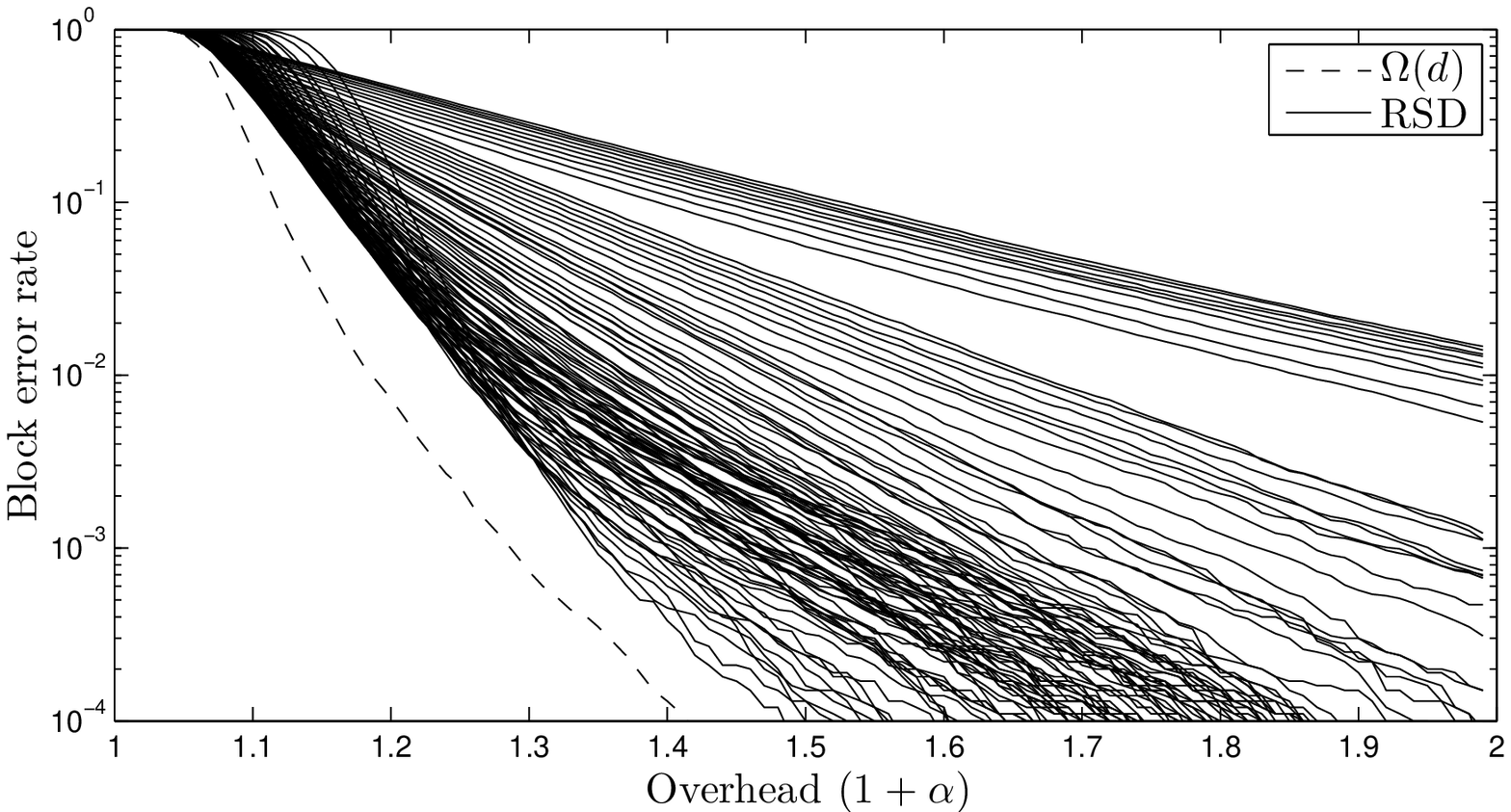}
\caption{Comparison of optimized $\Omega(d)$ with all RSDs in Table \ref{tab:rsd}.}
\label{fig:mass_comp}
\end{figure}

Fig. \ref{fig:avg_comp} shows a comparison of the optimized degree distributions with respect to average overhead for different values of $k$. It clearly shows that the proposed design with a decreasing ripple size outperforms the other designs on average. Fig. \ref{fig:1024_comp} shows how the different designs perform at fixed overheads with respect to block error rate at $k=1024$ for $100000$ iterations. This figure shows that the gain on average is not achieved through compromising the performance at high overheads. At any overhead, the proposed design provides a clear improvement. Finally, Fig. \ref{fig:mass_comp} shows a plot of the performance of all parameter pairs in Table \ref{tab:rsd} for the RSD, compared to the single optimized version of $\Omega(d)$. This shows that the optimized $\Omega(d)$ outperforms all the simulated RSDs at any fixed overhead. Relative results similar to what is shown in figures \ref{fig:1024_comp} and \ref{fig:mass_comp} are observed at $k$ equal to $256$, $512$ and $2048$. 

\section{Conclusions}\label{sec:conclusion}
In this paper we have analyzed the sources of redundancy in LT codes. We arrived at the conclusion, that the probability of a symbol being redundant is a much faster increasing function of the ripple size when the symbol degree is high compared to when it is low. This means that the price of maintaining a high ripple size increases during the decoding process, since high degree symbols are utilized late in the decoding process. Motivated by this result, we proposed a design with a decreasing ripple size, as opposed to the original strategy of keeping it constant. A simple design procedure, which can provide any achievable target ripple size evolution was presented and the resulting degree distributions evaluated. The results show a significant performance increase compared to state of the art degree distributions, both with respect to average overhead and block error rate at any fixed overhead.



\bibliographystyle{ieeetr}
\bibliography{bibliography}

\end{document}